\documentclass[useAMS,usenatbib]{mn2e}
\usepackage{graphicx}

\newcommand{\ps}{\mbox{$P_{\rm s}$}}
\newcommand{\pb}{\mbox{$P_{\rm orb}$}}
\newcommand{\rn}{\mbox{$R_{\rm m}$}}
\newcommand{\rc}{\mbox{$R_{\rm c}$}}

\newcommand\lsim{\lower0.5ex\hbox{$\; \buildrel < \over \sim \;$}}
\newcommand\gsim{\lower0.5ex\hbox{$\; \buildrel > \over \sim \;$}}

\title[Exploration of Spin-down Rate of Neutron Star in High Mass X-ray Binaries]{Exploration of Spin-down Rate of Neutron Star in High Mass X-ray Binaries}

\author[Hai-Lang Dai£¬Xi-Wei Liu and Xiang-Dong Li]{Hai-Lang Dai$^{1}$\thanks{E-mail:
hldai@nju.edu.cn}, Xi-Wei Liu$^{2}$ and Xiang-Dong Li$^{3,4}$\\
$^{1}$Department of Physics, Faculty of Science,
JiangSu University, Zhenjiang 212013, China\\
$^{2}$College of Science, Huazhong Agricultural University, Wuhan
430070, China\\
$^{3}$Department of Astronomy, Nanjing University, Nanjing 210046,
China\\
$^{4}$Key Laboratory of Modern Astronomy and Astrophysics (Nanjing
University), Ministry of Education, Nanjing 210046, China }

\begin{document}


\pagerange{\pageref{firstpage}--\pageref{lastpage}}

\maketitle

\label{firstpage}

\begin{abstract}

We use the evolutionary population synthesis method to investigate
the statistical properties of the wind-fed neutron star (NS) compact
($P_{\rm orb}<10$ days) high-mass X-ray binaries (HMXBs) in our
Galaxy, based on different spin-down models. We find that the
spin-down rate in the supersonic propeller phase given \textbf{by
assuming that the surrounding material is treated as forming a
quasi-static atmosphere} or  \textbf{by assuming that the
characteristic velocity of matter and the typical Alfv$\acute{e}$n
velocity of material in the magnetospheric boundary layer are
comparable to the sound speed in the external medium} is too low to
produce the observed number of compact HMXBs. We also find that the
models suggested  \textbf{by assuming that the infalling material is
ejected with the corotation velocity at the magnetospheric radius
when the magnetospheric radius is larger than the corotation radius}
and  \textbf{by simple integration of the magnetic torque over the
magnetosphere} with a larger spin-down rate than that given by
\citet{dav81} or \citet{is75} can predict a reasonable number of
observed wind-fed NS compact HMXBs. Our calculated results indicate
that subsonic propeller phase may not exist at all by comparing with
the observed particular distributions of wind-fed NS compact HMXBs
in the $P_s-P_{orb}$ diagram. However, the spin-down rate suggested
by \citet{wan85,dai06,jia05} and that given by \citet{dav73} both
seem reasonable to produce the observed distribution of wind-fed NS
compact HMXBs in the $P_s-P_{orb}$ diagram. We cannot find which
spin-down rate seems more reasonable from our calculations.
\end{abstract}


\section{Introduction}

\textbf{HMXBs are composed of a compact object that orbits a massive
($>10~M_{\odot}$) donor star. The X-ray emission is due to the
accretion of matter from the donor star onto the compact companion
(black hole or neutron star).} In most cases, the donor stars do not
fill their Roche lobes and the compact objects accrete from the
stellar wind. Canonically, HMXBs can be roughly divided into two
groups: supergiant binaries and Be/X-ray binaries. In the supergiant
systems, either Roche-lobe or stellar wind accretion occurs, while
in the Be systems commonly only the latter process takes place since
the Be star is well inside its Roche lobe \citep{tau06}. Be/X-ray
binaries are both transient and persistent X-ray sources. Transient
systems are characterized by type II outburst during which their
flux increases by a factor of 10-10$^4$ over the quiescent level. On
the other side, persistent Be/X-ray binaries show a rather flat
lightcurve, lower luminosity, longer spin and orbital periods
\citep{rei11}.

Supergiant Fast X-ray Transients (SFXT) unveiled in the last few
years mainly thank to INTEGRAL observations of the Galactic plane
are a new sub-class of supergiant HMXBs  that display extreme
flaring behaviour on short ($\sim$ hour) timescales
\citep{sgu05,sgu06,neg06}. They host a massive OB supergiant star as
identified by optical spectroscopy. The compact object is generally
assumed to be a NS because of the broad band X-ray spectral shape
(0.2$-$100 keV) strongly resembling those of accreting X-ray pulsars
in classical HMXBs \textbf{\citep{whi95}}. A distinctive property of
SFXTs is the high dynamic range, spanning three to five orders of
magnitude, with sudden increases in luminosity from $10^{32}$ erg
$s^{-1}$ up to the flare peak luminosity \citep[ e.g.][]{in05}.
There are currently 10 confirmed and about as many candidate SFXTs
\citep{sid11}.

The INTEGRAL observatory appears to have discovered a class of
compact high mass X-ray binaries \textbf{which are a new class of
$\gamma-$ray sources for which a mechanism is presented by
\citet{bed09}}, \textbf{i.e., accreting neutron stars inside binary
systems}. These newly discovered massive binaries are compact with
orbital periods between a few to several days. Some contain
relatively slowly rotating NSs that may allow the material to
penetrate the inner NS magnetosphere.

The spin-down rate of NS in a \textbf{wind-fed NS} HMXB has been
investigated by many authors. The rate of loss of angular momentum
of a NS, proposed by \citet{is75}, \citet{fab75}, \citet{wic75}, is
obtained by assuming that the characteristic velocity of matter
carrying off the required energy and the typical Alfv$\acute{e}$n
velocity of material in the magnetospheric boundary layer are
comparable to the sound speed in the external medium. The rate,
proposed by \citet{dav73}, \citet{kun76}, \citet{van78} is obtained
by simple integration of the magnetic torque over the magnetosphere
and equating this to an angular momentum loss from the star by
assuming that the field lines at the magnetosphere are swept back
through an angle of 45$^{o}$. The most rapid spin-down rate,
proposed by \citet{sha75}, \citet{lip76}, \citet{hol78}, is obtained
by treating the particle striking the magnetosphere as independent
particles, and by assuming that they are all accelerated to a
characteristic velocity, comparable to the rotational speed of the
magnetosphere which is much larger than the sound speed in the
external medium. The comprehensive picture by assuming that the
surrounding material is treated as forming a quasi-static atmosphere
through which energy is transported was first drawn by
\citet{dav81}, whose model passes through four distinct phases as a
NS slows down. \citet{mor03} suggested that two parameters can
classify many proposed propeller spin-down models. Many authors have
derived the spin-down rate of propeller phase by assuming that the
infalling material is ejected with the corotation velocity at the
magnetospheric radius when the magnetospheric radius is larger than
the corotation radius \citep{wan85, dai06, jia05}. \textbf{Besides,
a large number of authors \citep[see, e.g.,][]{spr93, rap04, dan10,
dan12} have investigated the spin-down rate of NS accreting from a
disk in a NS HMXB. \citet{rom03} have studied accretion onto a star
in the propeller regime by magnetohydrodynamic simulations.}

\textbf{Though the propeller effect has been investigated
extensively, there still remain large uncertainties about the
efficiency of angular momentum loss during the propeller regime
\citep[see, e.g.,][]{pri72, ikh01}.} The investigations mentioned
above were usually either theoretical or numerical. To better
understand the spin-down mechanism of a NS, \textbf{we} should use
an evolutionary population synthesis which incorporates the
evolution of a neutron star's spin. In the present paper, we
describe a population synthesis study of the spin evolution of a NS
in a massive binary. Obviously, it is very difficult to provide more
stringent constrains on these spin-down models from theory. However,
we can give more constrains on the spin-down rate by comparing the
calculated results based on the evolutionary population synthesis
method by adopting different spin-down models with the observed
population of compact HMXBs. We describe the theoretical
considerations in \S 2. In \S 3, we present the calculated results.
Finally, we present a brief discussion and conclusions in \S 4.

\section{Model}
\subsection{Spin Evolution}
We consider a binary system consisting of a $1.4 M_{\odot}$
magnetized NS and a massive donor star. The spin-down evolution of a
NS in a binary system has been investigated by many authors.
Generally speaking, the spin-down evolution of a NS before steady
accretion occurs contains two phases: the pulsar phase and the
propeller phase, as briefly presented below.

\noindent{\em Case 1: the pulsar phase}

Following its birth in a supernova explosion, the NS in a binary
system first appears as a radio pulsar with a short spin period, if
its radiation is strong enough to expel the wind material outside
the Bondi accretion radius $r_{\rm G}=2GM/v_{\infty}^2$
\citep{bon44}($G$, the gravitational constant, $M$, the mass of the
NS and $v_{\infty}=10^8v_8$ cm s$^{-1}$, the relative wind velocity
at the neutron star's orbit), or the radius of the light cylinder,
$r_{\rm lc}=c\ps/2\pi$. Magnetic dipole radiation and/or energetic
particle emission result in the spin-down of a NS:

\begin{equation}
I\dot{\Omega}_{\rm s}=-\frac{2}{3}\frac{\mu^2\Omega_s^3}{c^3},
\end{equation}
where $I$, the moment of inertia, $\mu=10^{30}\mu_{30}$ G cm$^3$,
the magnetic dipole moment, and $\Omega_{\rm s}$, the angular
velocity of the neutron star, respectively.

The pulsar phase will end either when the wind plasma penetrates
inside the light cylinder radius $r_{\rm lc}$ or when the pressure
gradient dominates at large radius. The corresponding transitional
spin period $P_{\rm ab}$ derived by balancing radiation pressure
from the pulsar with the stellar wind ram pressure at $r_{\rm lc}$
and $P_{\rm ac}$ obtained as the outer boundary $R_{\rm a}$ of the
envelope approaches $r_{\rm G}$ can be described,

\begin{equation}
P_{\rm ab} \simeq 0.8 \mu_{30}^{1/3} \dot{M}_{15}^{-1/6}
(M/M_{\odot})^{1/3}v_8^{-5/6}\,{\rm s},
\end{equation}

\begin{equation}
P_{\rm ac}\simeq
1.2\dot{M}_{15}^{-1/4}\mu_{30}^{1/2}v_8^{-1/2}\,{\rm s}
\end{equation}
\citep{dav81} ,where $\dot{M}=10^{15}\dot{M}_{15}$
 g s$^{-1}$, the accretion rate of the NS.

 \noindent{\em Case 2: the propeller phase}

The propeller phase begins when the pulsar phase breaks down if the
magnetospheric radius $\rn=[\mu^4/(2GM\dot{M}^2)]^{1/7}$ is larger
than the corotation radius $\rc=(GM/\Omega_{\rm s}^2)^{1/3}$. The
angular momentum of the NS is taken away from the NS surface when
the infalling plasma is ejected outward because the centrifugal
barrier inhibits further accretion. Although many authors
\citep[e.g.][]{pri72,is75,dav81,wan85,ikh01,boz08,sha12} have
investigated the propeller effect extensively, the efficiency of
angular momentum loss during the propeller phase is still uncertain.

\citet{dav81} suggested that the propeller phase can be divided into
two subphases: supersonic propeller phase and subsonic propeller
phase.

(a) The NS will enter the supersonic propeller phase when the
angular velocity, $\Omega_s$, of the neutron star is large enough so
that $r_c\Omega_s\gg c_s(r_c) $, where $r_c$ is the inner boundary,
$c_s(r_c)$ is the sound speed at the radius of $r_c$. The spin-down
torque is

\begin{equation}
N=-8\times10^{31}\dot{M_{15}}v_{8}^2\Omega_{\rm s}^{-1}{\rm g
cm^{2}s^{-2}}
\end{equation}
\citep[case c of][]{dav81}. The typical spin-down time-scale is
\begin{equation}
\tau= 1.6\times
10^7\dot{M}_{15}^{-1}v_{8}^{-2}I_{45}P_{0}^{-2}\,{\rm yr}.
\end{equation}
The spin-down of supersonic propeller phase process ends until $\ps$
reaches the equilibrium spin period
\begin{equation}
P_{\rm eq}= 23\mu_{30}^{2/3}\dot{M}_{15}^{-1/3}v_{8}^{-2/3}\,{\rm
s}.
\end{equation}

(b)\citet{dav81} suggested that accretion is unlikely to take place
unless the material outside the magnetosphere can cool. Therefore,
the NS spins at a subsonic speed and continues to lose rotational
energy. The spin-down torque of subsonic propeller phase is

\begin{equation}
N=-1.2\times10^{36}\mu_{30}^2(M/M_{\odot})^{-1}P_{0}^{-3}\Omega_{\rm
s}^{-1}{\rm g cm^{2}s^{-2}}
\end{equation}
\citep[case d of][]{dav81}. The typical spin-down time-scale is
\begin{equation}
\tau\simeq 10^3\mu_{30}^{-2}(M/M_{\odot})P_{0}I_{45}\,{\rm yr}.
\end{equation}
The spin-down of subsonic propeller phase process ends until $\ps$
reaches $P_{br}$
\begin{equation}
P_{\rm br}=
60\mu_{30}^{16/21}\dot{M}_{15}^{-5/7}(M/M_{\odot})^{-4/21}\,{\rm s}.
\end{equation}

Some authors proposed a different spin-down model for the propeller
phase \citep{wan85,dai06,jia05,ikh01}. They assumed that the plasma
is accelerated outward with the corotation velocity at $\rn$, and
the spin-down torque is
\begin{equation}
N=I\dot{\Omega}_{\rm s}=-\dot{M}R_{\rm m}^2\Omega_{\rm s}.
\end{equation}
The typical spin-down time-scale $\tau=|\Omega_{\rm
s}/\dot{\Omega}_{\rm s}|$ can be estimated to be
\begin{equation}
\tau\simeq 2.2\times
10^4\mu_{30}^{-8/7}\dot{M}_{15}^{-3/7}(M/M_{\odot})^{2/7}I_{45}\,{\rm
yr}
\end{equation}
\citep{dai06}. The process of spin-down ceases when $\ps$ reaches
the equilibrium spin period
\begin{equation}
P_{\rm eq}\simeq
17\mu_{30}^{6/7}\dot{M}_{15}^{-3/7}(M/M_{\odot})^{-5/7}\,{\rm s}.
\end{equation}

\citet{mor03} suggested that many proposed propeller spin-down
models can be classified by two parameters. In these models, the
spin-down torque is
\begin{equation}
N=I\dot{\Omega}_{\rm s}=-\dot{M}R_{\rm m}v_{m}\mathcal{M}^{\gamma},
\end{equation}
where $\mathcal{M}$ is the Mach number defined as the ratio of
incoming medium velocity to NS spin-velocity at the magnetosphere
boundary : $\mathcal{M}\equiv R_m\Omega/v_m$. Proposed propeller
models have $\gamma$ with the value of -1, 0, 1 and 2.

Steady wind accretion onto the surface of the NS occurs when
$P>P_{\rm eq}$. However, the present spin periods of wind-fed X-ray
pulsars are not significantly different from the $P_{\rm eq}$. So we
stop the calculations when either $P_{\rm eq}$ is reached within the
main sequence lifetime or the optical star evolves off the main
sequence. In the present paper, the narrow HMXBs with Roche-lobe
overflow are never considered because the accreting material is most
likely to come from an accretion disk. \textbf{For wind-fed systems
like Vela X$-$1, numerical calculations
\citep[e.g.][]{fry88,mat92,anz95,ruf99} suggest that there are no
significant angular momentum transfer onto the neutron star when
radially expanding wind matter is transferred onto the neutron star.
This may lead to only small deviation from the instantaneous
(equilibrium if reached) spin periods when the accretion phase
starts. {A random walk in their spin frequencies with alternating
spin-up and spin-down \citep{bil97} is shown by \em CGRO/BATSE}
observations .}

\subsection{Evolution of the mass-flow rate onto NS}

We used the evolutionary population synthesis code developed by
\citet{hur00,hur02} to explore the spin-down rate of a NS in a
binary system. The evolution of single stars with binary-star
interactions, such as mass accretion, mass transfer, common-envelope
(CE) evolution, collisions, supernova kicks, angular momentum loss
mechanisms and tidal friction, is included in this code. The
parameters we adopted are mostly the same as those described by
\citet{hur02}. The primary-mass ($M_{\rm 1}$) distribution is the
initial mass function of \citet{Kro93}. A uniform distribution of
the mass ratio $0<q\equiv M_{\rm 2}/M_{\rm 1}\leq 1$ is taken
between 0 and 1 for the secondary star (of mass $M_{\rm 2}$). For
the binary separation a, we take a uniform distribution in
\textbf{{\rm $lna$} (natural logarithm of a).} We assume that one
binary with $M_{\rm 1} \geq8M_{\odot}$ is born in the Galaxy per
year, which gives the star formation rate $S = 7.6085$ yr$^{-1}$.
During the SN explosions, a kick velocity with the Maxwellian
distribution is imparted on a NS with a mean of 265 km s$^{-1}$
\citep{hobb05}. \citet{hur02} presented specially the treatment of
Roche-lobe overflow (RLOF) mass transfer in the primordial binary
and the stability criterion of mass transfer is described briefly
here. According to whether the primary stays in thermal equilibrium
when it loses mass, and the radius of the primary increases faster
than the Roche-lobe, mass transfer through RLOF takes place on
either a thermal, nuclear, or dynamical time-scale. Stars with deep
surface convective zones---for instance, giants or naked helium
giants---will enter a CE evolution because they are generally
unstable to dynamical-timescale mass loss. Eddington accretion rate
limits the stable mass accretion rate of the secondary star. We
haven't considered the situation that the secondary may be spun up
and become a Be star as it accretes enough mass because the origin
of Be phenomena is still unclear and it is hard to give a model of
the mass transfer processes in Be/X-ray binaries. The common envelop
efficiency parameter $\alpha$ was set at $\alpha=1$ as a typical
value and we varied it from 0.1 to 2 in our calculations
\citep{dew00, tau01}. The region of the parameter space from which
NS high-mass binaries form is shown in Fig. 1. We can also derived
other binary parameters, such as the surface temperature,
luminosities, and radii of the companion stars. The mass loss rates
from the companion stars and the mass flow rates onto the NSs can be
evaluated by using these parameters.

\begin{figure}
\centering
\includegraphics[width=84mm]{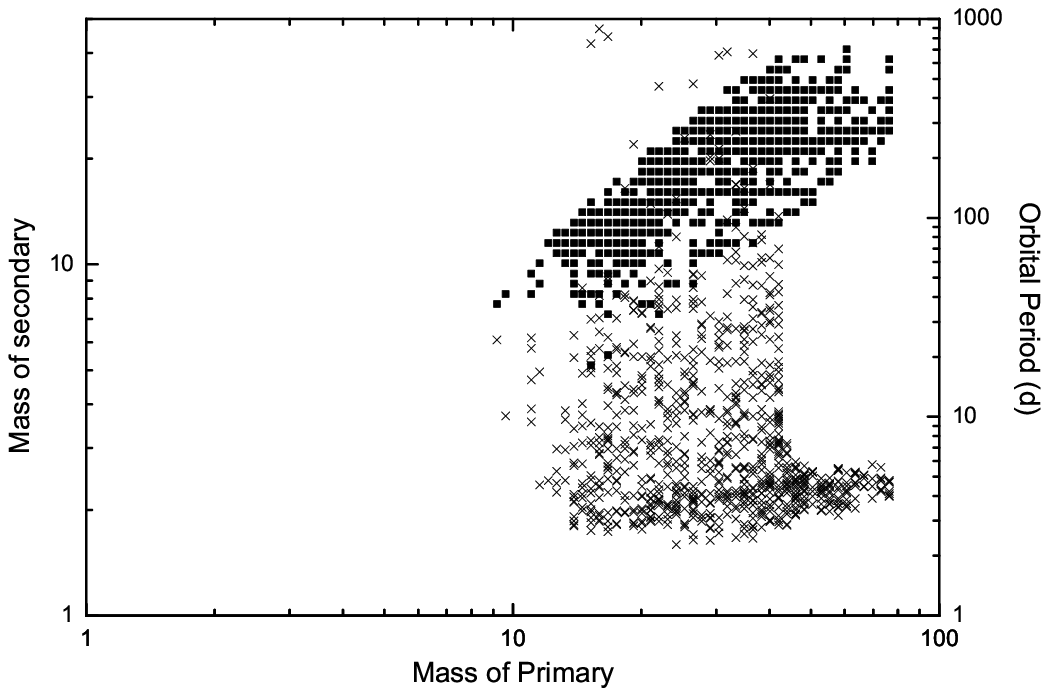}
\includegraphics[width=84mm]{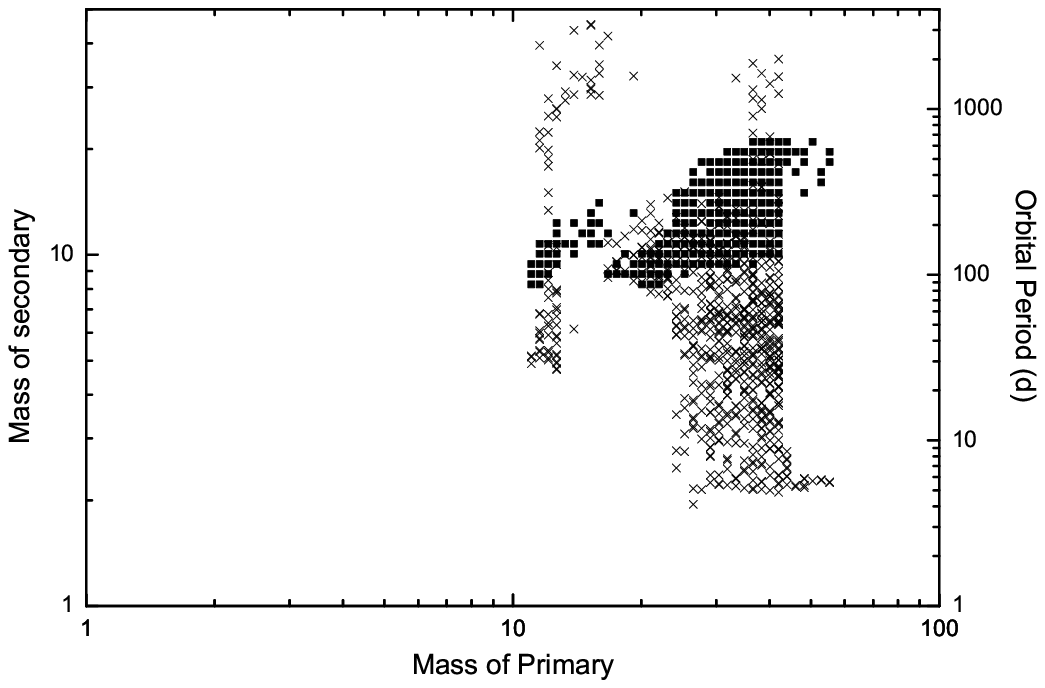}
   \caption{The region of parameter space from which NS high-mass
   binaries form. a) The mass transfer before primary star
   core-collapsing is dynamically stable; b) That mass transfer is
   dynamically unstable and thus the progenitor binary has evolved
   through a CE phase. Filled squares denote the mass distribution,
   while crosses denote the primary mass-orbital period
   distribution. The CE efficiency factor is adopted as a typical
   value $\alpha=1.0$.
    }
   \label{}
\end{figure}

To compare the observed properties of compact wind-fed NS HMXBs, we
have calculated the number of the compact wind-fed NS HMXBs with
$\pb<10$ days. The mass-loss rate $\dot{M}_2$ was presented by
\citet{nie90}:
\begin{equation}
-\dot{M}_2=9.6\times10^{-15}
R_2^{0.81}L_2^{1.24}M_2^{0.16}\,M_{\odot}{\rm yr}^{-1},
\end{equation}
where $R_2$ and $L_2$ are the radius and luminosity of the donor
star. We evaluate all the physical quantities in Eq.~(14) in solar
units. The wind density $\rho_{\rm w}$ at the orbit of the NS by
assuming that the stellar wind expands isotropically at a speed of
$v_{\rm w}$,  is
\begin{equation}
\rho_{\rm w}=-\dot{M}_2/(4\pi a^2v_{\rm w}),
\end{equation}
and the mass infalling rate onto NS is roughly described by

\begin{equation}
\dot{M}=\pi r_{\rm G}^2\rho_{\rm w}v_{\infty}
\end{equation}
\citep{bon44}.

\section{Results}
We calculated the evolution of spin and the statistical properties
for compact wind-fed X-ray pulsars in NS binary systems based on the
theoretical models presented in \S 2. For the initial NS magnetic
fields $B$ we assumed that $\log B$ is distributed normally with a
mean of 12.5 \textbf{(the typical magnetic fields of such pulsars
are $3\times 10^{12}\rm{G}$)} and a standard deviation of 0.3. No
field decay was considered. We stopped our calculations for the spin
evolution when either $\ps$ reaches $P_{\rm eq}$/$P_{br}$ (depending
on whether the subsonic propeller phase exists) or the companion
star began to evolve off the MS , for the reasons described
\textbf{above}.

We adopt a variety of models (see Table 1), each with different
assumptions for the spin-down rate and parameters that govern the
evolutions in the calculations. In Table 1, DP81, WR85, MR03
represent the spin-down models described by \citet{dav81},
\citet{wan85}, and \citet{mor03} respectively. NO in Table 1
describes the situation when the subsonic propeller phase doesn't
exist. We consider the only situations when $\gamma=-1$ and
$\gamma=0$ for the spin-down model described by \citet{mor03}
because the spin-down model corresponds to WR85 and DP81 when
$\gamma=1$ and $\gamma=2$ . Figure 2 shows evolution of the NS spin
period (with the same initial parameters of the binary) in a binary
system for different models, which indicates that it may induce a
longer spin period of NS when the subsonic phase exists.

\begin{table}
\begin{center}
\caption{Model parameters for binary population synthesis. The items
of ``sup'' and ``sub'' denote the adopted NS spindown models for the
supersonic propeller and subsonic propeller phases, respectively.
$\dot{M}=1$ or 3 denotes the wind mass-loss rates assuming to be 1
or 3 times of the standard prescription we adopted in \S 2,
respectively. $v_8$ is the wind velocity in unit of $10^8$ cm.}
\begin{tabular}{clrrrrrrrrrr}
\hline\hline \multicolumn{1}{c}{Model} & sup & sub &
$\alpha$ & $\dot{M}$ & $v_8$  \\
\hline
A1 &DP81 &DP81 &1    &1 &1\\
A2 &DP81 &NO   &1    &1 &1\\
A3 &DP81 &NO   &0.5  &1 &1 \\
A4 &DP81 &NO   &1    &3 &1 \\
A5 &DP81 &NO   &1    &1 &2 \\
B1 &WR85 &DP81 &1    &1 &1 \\
B2 &WR85 &DP81 &0.5  &1 &1 \\
B3 &WR85 &DP81 &1    &3 &1 \\
B4 &WR85 &DP81 &1    &1 &2 \\
B5 &WR85 &NO   &0.5  &1 &1 \\
B6 &WR85 &NO   &1    &1 &1 \\
B7 &WR85 &NO   &0.5  &1 &2 \\
C1 &MR03($\gamma=-1$) &DP81   &0.5  &1 &1 \\
C2 &MR03($\gamma=0$)  &DP81   &0.5  &1 &1 \\
C3 &MR03($\gamma=-1$) &NO   &0.5  &1 &1 \\
C4 &MR03($\gamma=0$)  &NO   &0.5  &1 &1 \\
C5 &MR03($\gamma=-1$) &NO   &1    &0.5 &1 \\
C6 &MR03($\gamma=0$)  &NO   &1    &0.5 &1 \\

\hline
\end{tabular}


\end{center}
\end{table}

\begin{figure}
 \includegraphics[width=84mm]{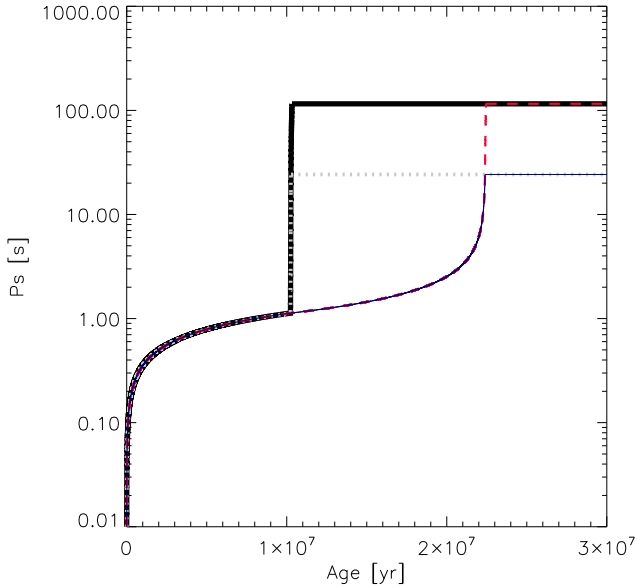}
   \caption{An evolutionary example of compact NS-HMXBs.
   Evolution of NS spin period in a binary system with the same initial parameters of the binary for model A1 (red dashed line),
   model A2 (thin blue line), model B1 (thick black line), and
   model B6 (thick gray dotted line). The evolution begins after the
   birth of the NS.
    }
   \label{}
\end{figure}

Table 2 summarizes the calculated numbers of compact
($P_{orb}<10$~days) wind-accreting NS HMXBs in our Galaxy for
different models (listed in Table 1). The observed compact
($P_{orb}<10$~days) NS HMXBs are listed in Table 3 \citep{liu06,
wan10, rei09,wan12, wan13,pea13,man12}. We find that the spin-down
rate in the supersonic propeller phase given by \citet{dav81} is too
low to produce the observed population of compact HMXBs no matter
whether the subsonic propeller exists or not. Our calculation shows
the similar conclusion for the spin-down model described by
\citet{mor03} when $\gamma=-1$. We also find that the model
suggested by \citet{wan85}, \citet{dai06} and \citet{jia05} with a
larger spin-down rate than that given by \citet{dav81} can predict a
reasonable number of observed wind-fed compact NS HMXBs no matter
whether the subsonic propeller phase exists or not. We can also
derive the similar conclusion for the spin-down model described by
\citet{mor03} when $\gamma=0$.

\begin{table*}
\begin{center}
\begin{minipage}{140mm}
\caption{Predicted present numbers in our Galaxy of Compact
($P_{orb}<10$ days) wind-accreting NS-HMXBs. }
\begin{tabular}{cccccccrrrrr}
\hline\hline \multicolumn{1}{c}{Model} & $A1$ & $A2$ &
$A3$ & $A4$& $A5$  &  $B1$\\
\hline
Number &$9.1\times 10^{-3}$ &$9.2\times 10^{-3}$ &$1.3\times 10^{-2}$ &$5.3\times 10^{-1}$ &0  &6.4 \\
\hline
Model &B2&B3 &B4 &B5 &B6 &B7 \\
\hline
Number &9.9  &1.9  &6.5 &10.3  &6.7 &13.6 \\
\hline
Model & C1  &  C2& C3 & C4 &C5 &C6\\
\hline
Number  &0 &4.8 &0 &4.9 &0 & 4.9\\

\hline
\end{tabular}


\end{minipage}
\end{center}
\end{table*}

\begin{table*}
\begin{center}
\begin{minipage}{140mm}
\caption{Observed compact ($P_{orb}<10$ days) NS-HMXBs.}
\begin{tabular}{cccccccrrrrr}
\hline\hline \multicolumn{1}{c}{Name} &$P_{orb}$(d)  &
$P_{pulse}$(s)& Name &
$P_{orb}$(d) & $P_{pulse}$(s)  \\
\hline
1WGA J0648.0-4419 &1.55  &13.1789 &4U 0900-40 &8.96 &283\\
RX J0648.1-4419 & & & Vela X-1 &  & \\

\hline
4U 1119-603 &2.09  &4.84& 4U1538-52  &3.73 &529  \\
Cen X-3 & &  &  & \\

\hline
IGR J16320-4751 & 8.96 & 1309 &4U 1700-37\footnote[1]{$M_X=2.44$, low mass black hole candidate?}  & 3.41 & \\
AX J1631.9-4752 &  &  &  &  \\

\hline
EXO 1722-363 & 9.74 &  413.9 & SAX J1802.7-2017 & 4.6 &139.61\\
IGR J17252-3616 &   &   &IGR J 18027-2016   &  \\

\hline
XTE J 1855-026 &  6.067 &361 &4U 1907+09  & 8.38 & 438\\
& & &H 1907+097 & \\

\hline 4U 1909+07 & 4.4  & 604.68  &4U 2206+543\footnote[2]{392 s
pulsation \citep{rag05}?} &9.57\footnote[3]{a new periodicity of
16.25 d recently suggested by
\citet{rei09}} & 5560\\
X1908+075 &   &  &3A 2206+543  & \\

\hline IGR J16493-4348 & 6.78 &1093 &IGR J16418-4532 & 3.74 &
1240\\

\hline IGR J17544-2619 & 4.9 & 71 & IGR J01583+6713 &
3-12\footnote[4]{a possible orbital period in the
range 3-12 days suggested by \citet{wan10}}&469 \\

\hline

\end{tabular}


\end{minipage}
\end{center}
\end{table*}

To compare the calculated results with observations of compact
wind-fed NS HMXBs, we show the distributions of those neutron star
binaries with $P\gsim P_{eq}$ and of the observed HMXBs in a
$P_s-P_{orb}$ diagram (see Figure 3 and 4). The relative numbers of
binary systems are indicated by the darkness of the shading.
Diamonds and pluses mark the supergiant wind-fed
 HMXBs and supergiant fast X-ray transients, respectively, and
 crosses are for Roche lobe overflow systems. The two joined triangles
 represent 4U 2206+54 for the two suggested orbital periods
 \citep{rei09}. The two joined asterisks represent IGR J01583+6713 for a
 possible orbital period in the range 3-12 days suggested by
 \citet{wan10}. Figure 3 presents the
 calculated results for those models when subsonic propeller phase
 exists. According to our calculations, we note that the spin periods
 are too large to compare with the observations no matter which values the
 parameters take when subsonic propeller phase exists. That is, no spin-down model
 can produce the observed distribution of compact wind-fed NS HMXBs in
 a $P_s-P_{orb}$ diagram when subsonic propeller phase exists.

 \begin{figure}

\centering
\centerline{
\includegraphics[width=42mm]{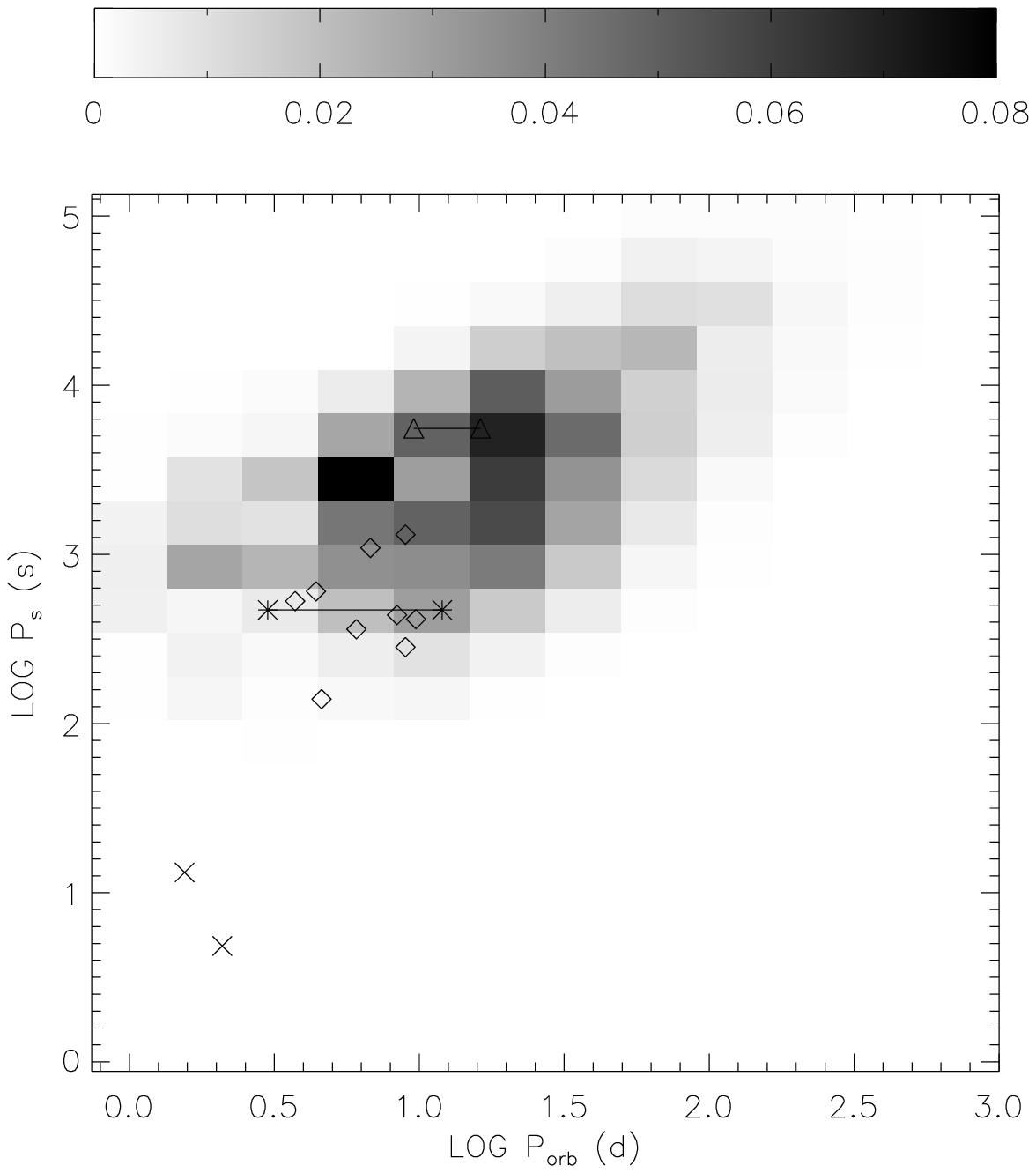}
\includegraphics[width=42mm]{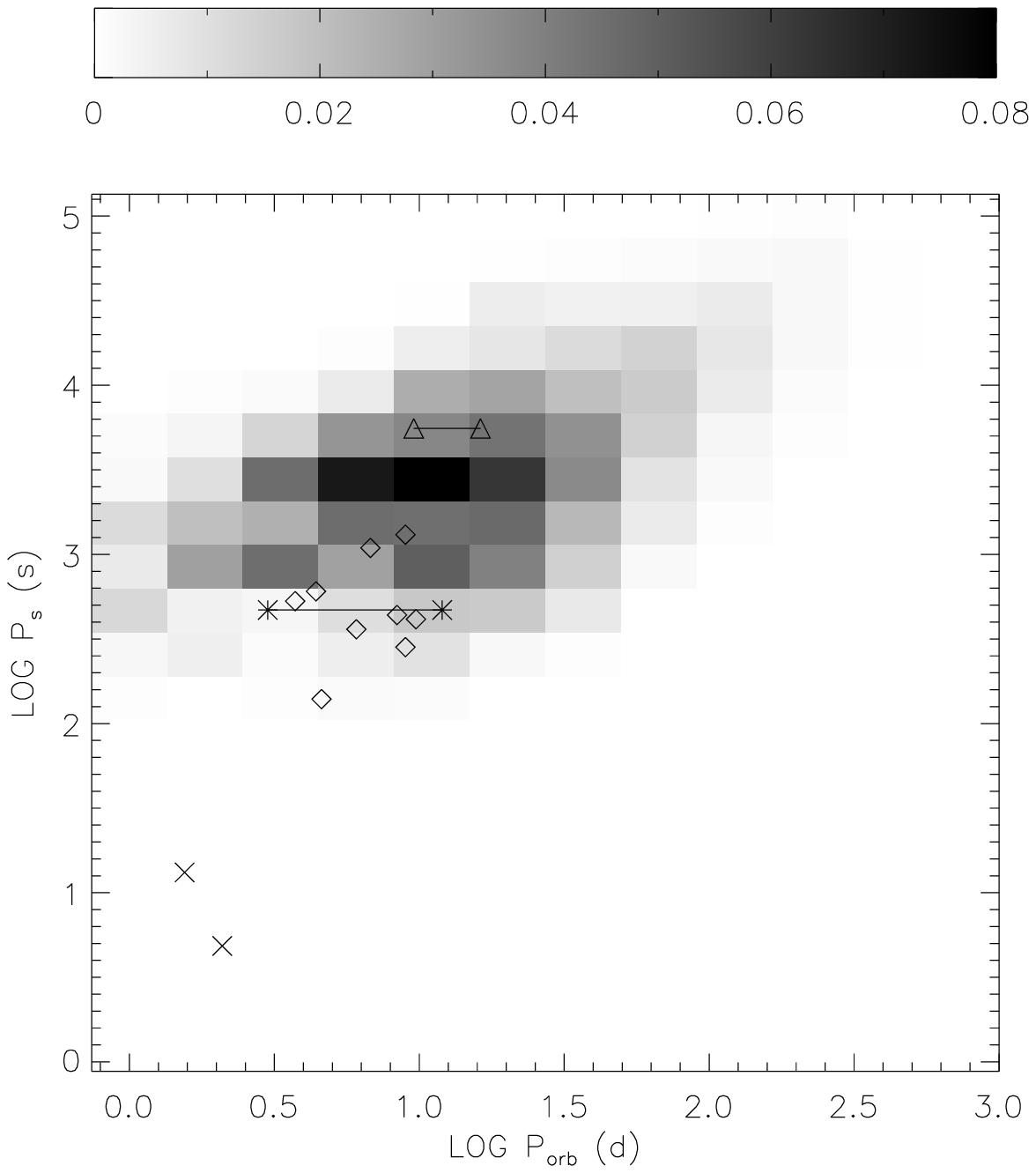}}
\centerline{
\includegraphics[width=42mm]{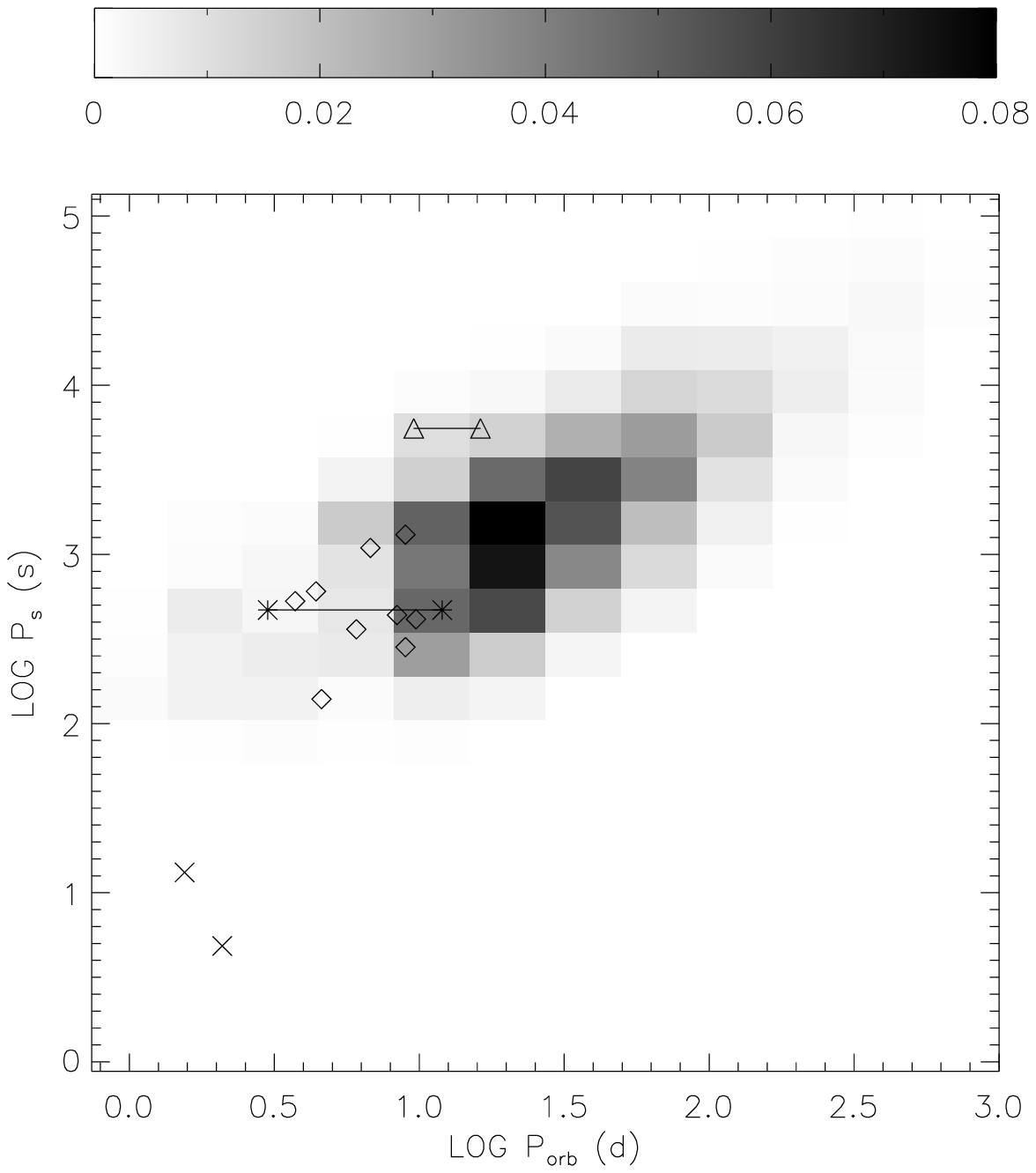}
\includegraphics[width=42mm]{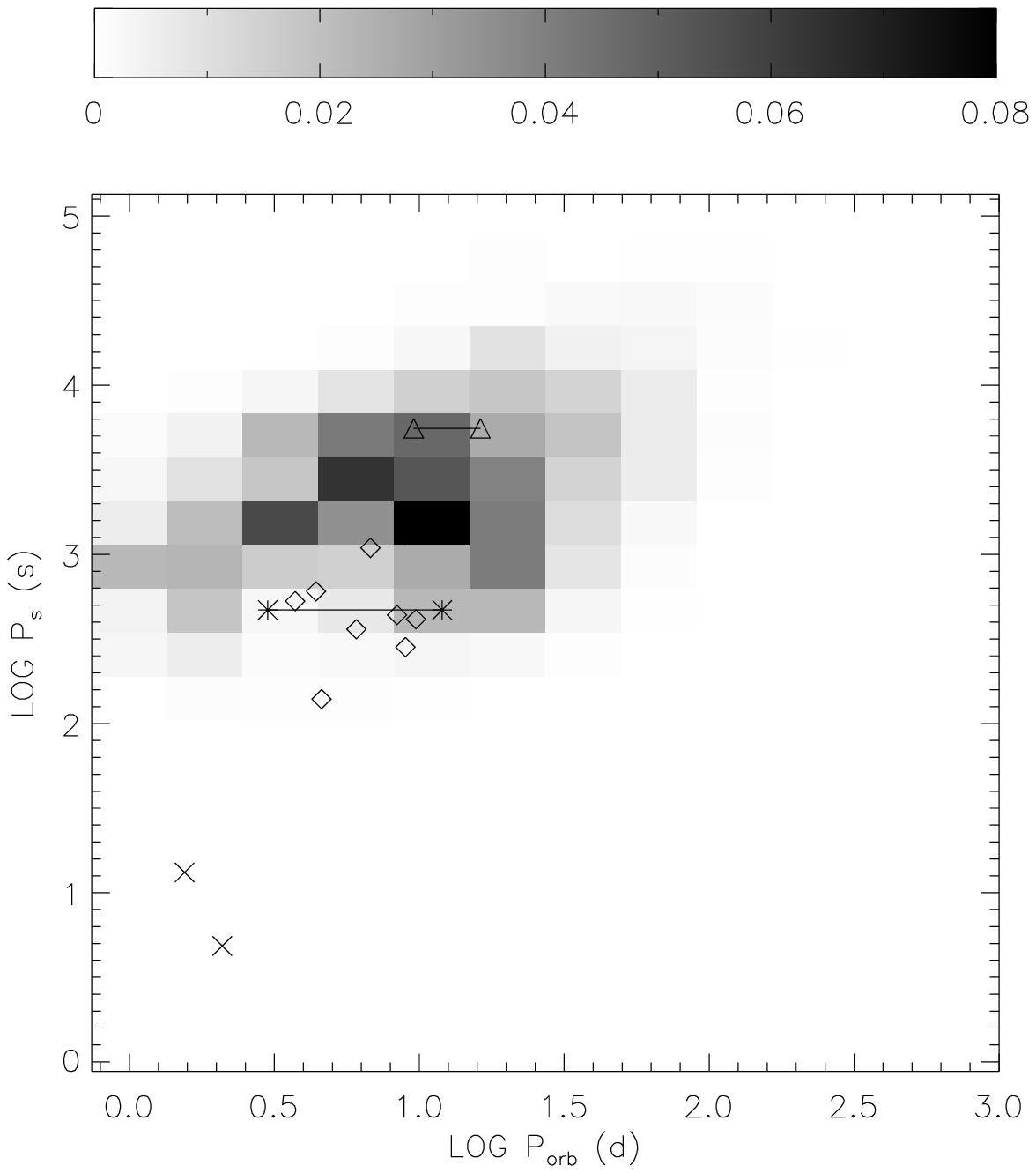}}
 \caption{The $P_s-P_{orb}$ distribution of wind-fed HMXBs.
 Diamonds and pluses mark the supergiant wind-fed
 HMXBs and supergiant fast X-ray transients, respectively, and
 crosses are for Roche lobe overflow systems. The two joined triangles
 represent 4U 2206+54 for the two suggested orbital periods
 \citep{rei09}. The two joined asterisks represent IGR J01583+6713 for a
 possible orbital period in the range 3-12 days suggested by
 \citet{wan10}. Top, model B1 (left) and model B2 (right); bottom,
 model B3 (left) and model C2 (right).
 }
\end{figure}


 Figure 4 shows the calculated results for all the spin-down models
 with different values of the parameter $\alpha$, $v_{8}$,
 and $\dot{M}$ when the subsonic propeller phase doesn't exist. If
 $v_8$ is increased from 1 to 2, the mass flow rates onto the
 neutron stars are lower by a factor of $\sim16$ in accordance with
 equations (15) and (16), further inducing longer equilibrium
 periods which can be seen clearly in Figure 4. The results also indicate
 that changes in the parameters $\alpha$ and $\dot{M}$ do not
 significantly influence the final outcomes, which is consistent
 with the observed distribution of compact wind-fed NS HMXBs in our Galaxy
 in a $P_s-P_{orb}$ diagram. So, we can conclude that the subsonic
 propeller phase may not exist at all from our calculated results plotted
 in Figure 3 and 4.

 \begin{figure}
\centering \centerline{
\includegraphics[width=42mm]{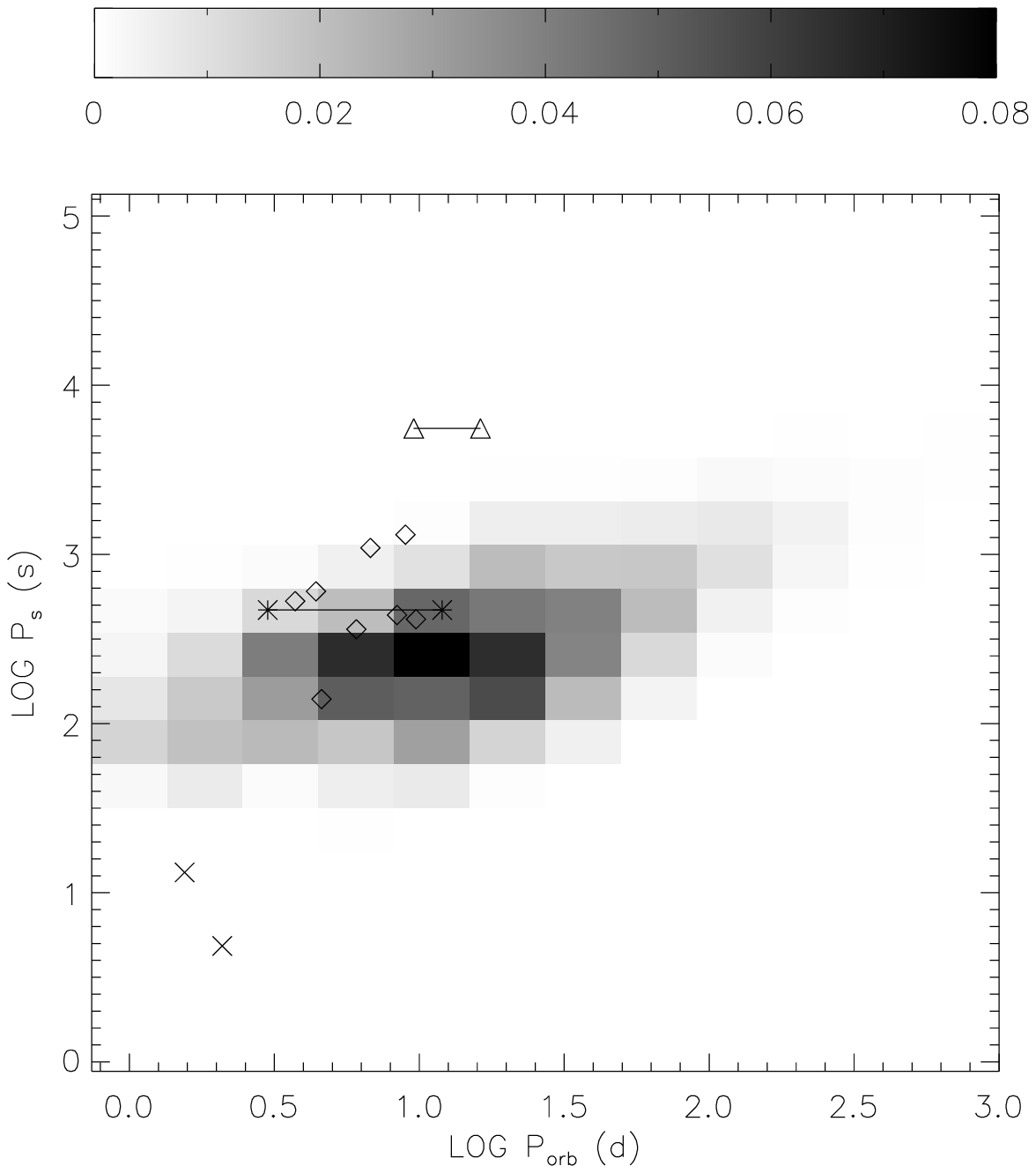}
\includegraphics[width=42mm]{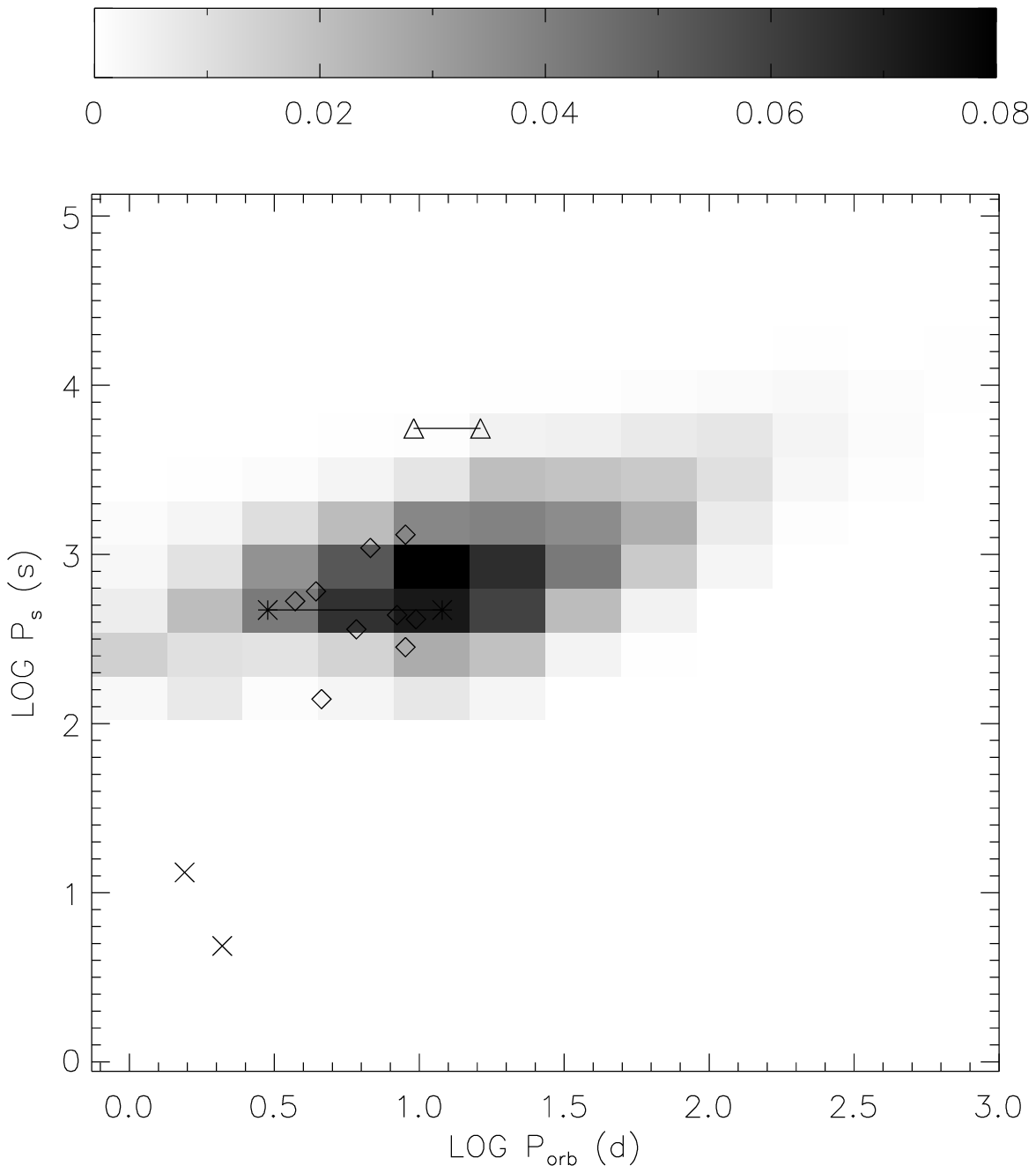}}

\centerline{
\includegraphics[width=42mm]{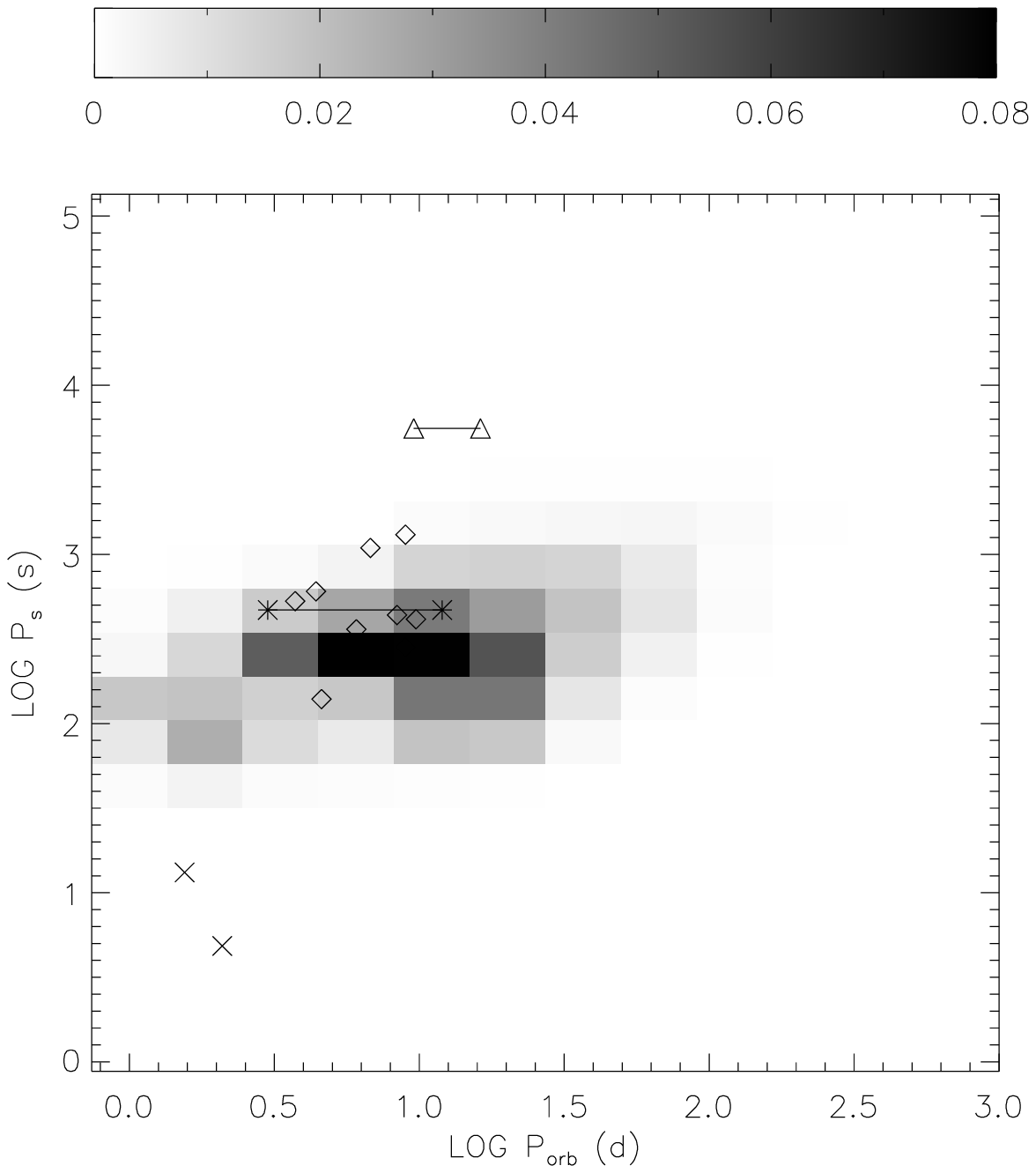}
\includegraphics[width=42mm]{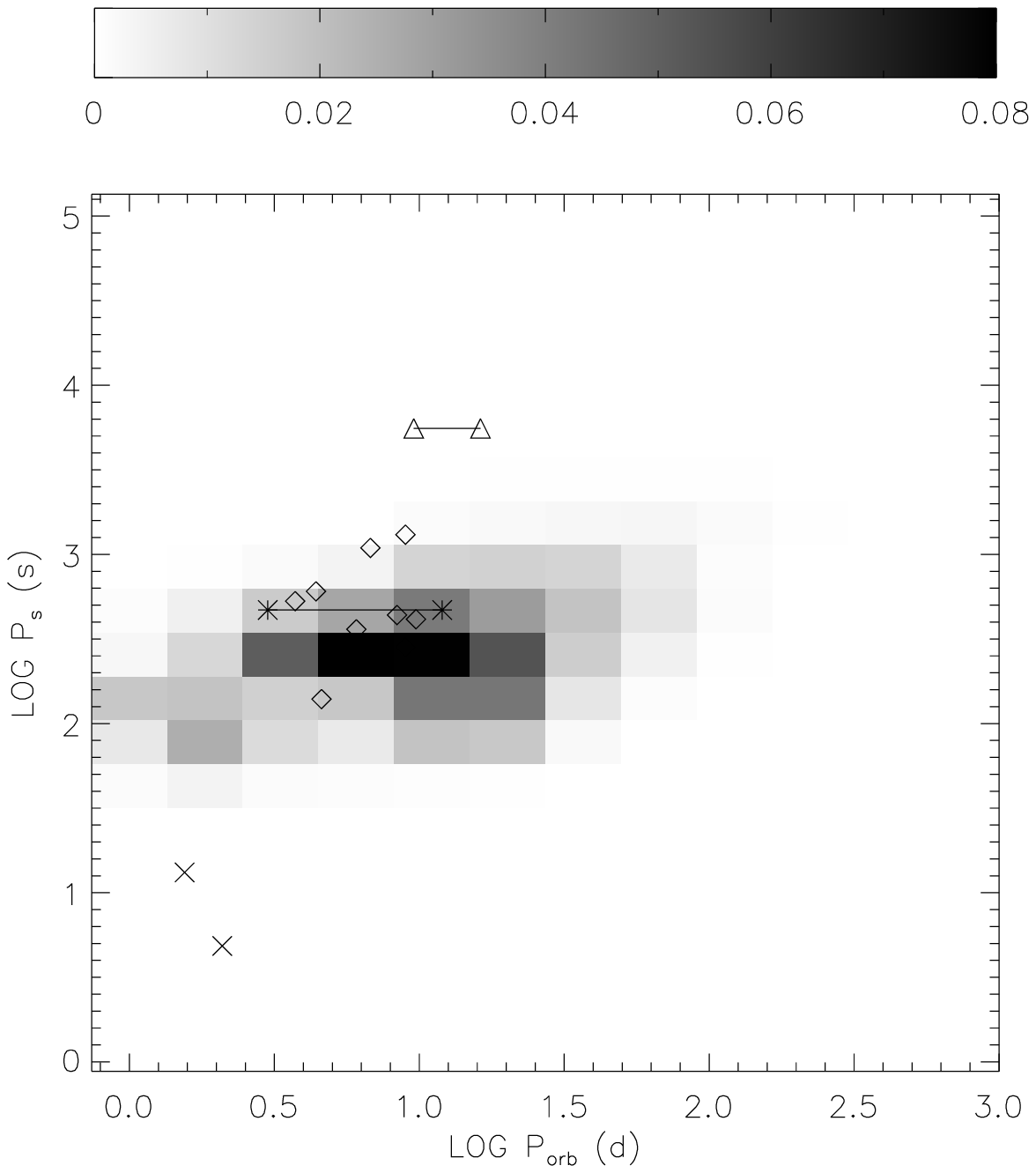}}


\caption{Same as Figure 3 but top, model B5 (left) and model B7
(right); bottom, model C4 (left) and model C6 (right).}
\end{figure}

 \section{Conclusions and Discussion}

We have calculated the evolution of spin and the statistical
properties for compact wind-fed X-ray pulsars in NS binary systems.
The numerical results presented in Table 2 show that the spin-down
rate in the supersonic propeller phase given by \citet{dav81} is too
low to produce the observed number of compact HMXBs no matter
whether the subsonic propeller phase exists or not. The same
conclusion can be derived for the spin-down rate in the supersonic
propeller phase given by \citet{is75} \citep[the case of $\gamma=-1$
in][]{mor03}. We also find that the spin-down model proposed by
\citet{wan85,dai06,jia05} can predict a reasonable number which is
consistent with the observations no matter whether the subsonic
propeller phase exists or not. The same conclusion can be inferred
for the spin-down rate in the supersonic propeller phase described
by \citet{dav73}  \citep[the case of $\gamma=0$ in][]{mor03}. In
order to investigate whether the subsonic propeller phase exists or
not, we compare our calculated results with the observed particular
distributions of compact supergiant HMXBs in the $P_s-P_{orb}$
diagram which has been described in \citet{dai06}. From Figure 3 and
4, we can conclude that the subsonic propeller phase may not exist
at all. The very long period, $P_{\rm s}=5560$ s, of 4U 2206+543,
may be explained by a accreting magnetar model which allows it to be
spun down efficiently by the propeller effect
\citep{ikh10,rei12,wan13}. However, the spin-down rate given by
\citet{dav81,dai06,jia05} and that given by \citet{dav73} in the
supersonic propeller phase both seem reasonable to produce the
observed distributions of compact supergiant HMXBs in the
$P_s-P_{orb}$ diagram. We cannot conclude which spin-down rate seems
more reasonable from our calculated results.

Our results are subject to some uncertainties. The different values
of parameters $\alpha$ and $\dot{M}$ do not have significant
influence on the final outcomes  \textbf{(e.g., the total number
changes only by a factor of 2 or 3 even when the parameter $\alpha$
varies from 0.1 to 2, what's more, the different values of parameter
$\alpha$ has little effect on the distributions in the $P_s-P_{orb}$
diagram)}. The changes in the parameter $v_8$ do not significantly
influence the calculated number of compact NS HMXBs while it can
induce longer equilibrium period with a larger wind velocity. This
indicates that there may exist no subsonic propeller phase further.
\citet{aer03} have suggested that the wind velocity of supergiant
increases with radius according to a $\beta$-law in some special
condition, however, our calculations indicate that it does not
significant affect our final results.  We have also investigated the
effect of the magnetic field. The models can produce more compact
HMXBs (about ten times more compact HMXBs are produced when the
initial NS magnetic field becomes ten times larger) and longer spin
and orbital periods with a larger initial magnetic field. The
results also support our conclusion that there may not exist
subsonic propeller phase further. However, some authors  have also
proposed that the magnetic field of NS may decay during the
evolution of the binary \citep{gep01,hol02,zha12}. If we assume all
the neutron star's magnetic field decays, our calculations indicate
that it can produce less number of compact HMXBs and shorter spin
and orbital periods while it has no significant effect on our
conclusion. A number of authors have also suggested that some
neutron stars receive low kick speeds of $\le50$ km $s^{-1}$ at
birth \citep{pfahl02,pod04,dew05}. If all the neutron stars were
born with such small kicks, our calculations indicate that there
should have been about 4-5 times more compact HMXBs produced.
However, it has no significant effect on the distribution in the
$P_s-P_{orb}$ diagram.

\section*{Acknowledgments}

We are grateful to an anonymous referee for a careful reading of the
manuscript and constructive comments that significantly improved
this paper. This work was supported by National Natural Science
Foundation of China under grant Nos. 11003006, 11133001 and
11333004, and Natural Science Foundation of Jiangsu province, China
under grant number BK2010338.

\bsp

\label{lastpage}

\end{document}